\documentclass[prd,aps,eqsecnum]{revtex4}
\usepackage{graphicx}
\newcommand{\be}{\begin{equation}}
\newcommand{\ee}{\end{equation}}
\newcommand{\Be}{\begin{eqnarray}}
\newcommand{\Ee}{\end{eqnarray}}
\newcommand{\f}{\frac}

\begin{document}
\title{Rotating wormhole and scalar perturbation}
\author{Sung-Won Kim\footnote{E-mail address: sungwon@mm.ewha.ac.kr}}
\affiliation{Department of Science Education, Ewha Women's
University, Seoul 120-750, Korea}

\date{\today}

\begin{abstract}

In this paper, we study the rotational wormhole and scalar
perturbation under the spacetime. We found the Schr\"odinger like
equation and consider the asymptotic solutions for the special
cases.

\end{abstract}
\maketitle
\section{Introduction}

The wormhole has the structure which is given by two
asymptotically flat regions and a bridge connecting two
regions\cite{MT88}. For the Lorentzian wormhole to be traversable,
it requires exotic matter which violates the known energy
conditions. To find the reasonable models, there had been studying
on the generalized models of the wormhole with other matters
and/or in various geometries. Among the models, the matter or wave
in the wormhole geometry and its effect such as radiation are very
interesting to us. The scalar field could be considered in the
wormhole geometry as the primary and auxiliary effects\cite{K00}.
Recently, the solution for the electrically charged case was also
found \cite{KL01}.

Among the models, the rotating wormhole is very interesting to us,
since Kerr black hole is the final stationary state of most black
holes. The Kerr metric has many insights in black hole physics as
the general black hole solution with angular momentum. Likewise,
the rotating wormhole is stationary and axially symmetric
generalization of the Morris-Thorne (MT) wormhole. The reason is
that it may be the most general extension of MT wormhole.
Teo\cite{T98} derived the rotating wormhole model from the
generally axially symmetric spacetime and have shown an example
with ergoregion and geodesics able to traverse wormhole without
encountering any exotic matter.

Meanwhile, scalar wave solutions in the wormhole
geometry\cite{KSB94,KMMS95} was in special wormhole model and the
transmission and reflection coefficients were found.  The
electromagnetic wave in wormhole geometry is recently
discussed\cite{BH00a} along the method of scalar field case. These
wave equations in wormhole geometry draws attention to the
research on radiation and wave.

In the recent paper, we found the general form of the
gravitational perturbation of the traversable wormhole\cite{K04},
which will be a key to extend the wormhole physics into the
problems similar to those relating to gravitational wave of black
holes. The main idea and resultant equation is similar to
Regge-Wheeler equation\cite{RW57} for black hole perturbation.

In this paper, we studied the rotating wormhole and examined a
couple of models to see their properties and geometric structures.
We also found the equation of the scalar perturbation for the
special example, rigid rotating wormhole. Here we adopt the
geometrical unit, {\it i.e.}, $G=c=\hbar=1$.

\begin{figure}
\begin{center}
\includegraphics[width=10cm]{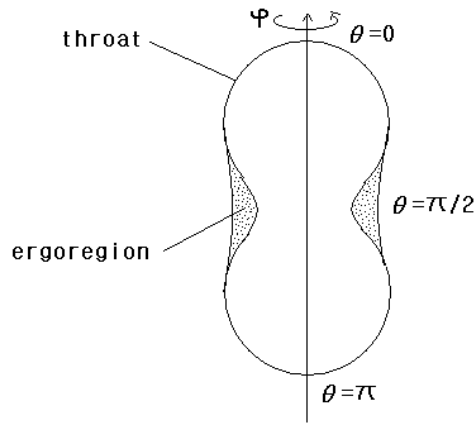}
\caption{Plot of the cross-sectional schematic of the rotating
wormhole throat\cite{T98}}
\end{center}
\end{figure}

\section{Rotating wormhole}

The spacetime in which we are interested will be stationary and
axially symmetric. The most general stationary and axisymmetric
metric can be written as \be ds^2 = g_{tt}dt^2 + 2 g_{t\phi}dt
d\phi + g_{\phi\phi}d\phi^2 + g_{ij}dx^idx^j, \ee where the
indices $i,j=1,2$. Teo\cite{T98} used the freedom to cast the
metric into spherical polar coordinates by setting
$g_{22}=g_{\phi\phi}/\sin^2x^2$\cite{T71} and found the metric of
the rotating wormhole as: \Be ds^2 &=& - N^2 dt^2 + e^\mu dr^2+
r^2K^2 d\theta^2 + r^2K^2\sin^2\theta [d\phi^2 - \Omega dt]^2  \\
&=& - N^2 dt^2 + \f{dr^2}{1-b(r)/r} + r^2K^2 d\theta^2 +
r^2K^2\sin^2\theta [d\phi^2 - \Omega dt]^2, \nonumber\
\label{eq:rot} \Ee where $\Omega$ is the angular velocity
$d\phi/dt$ acquired by a particle that falls freely from infinity
to the point $(r,\theta)$, and which gives rise to the well-known
dragging of inertial frames or Lense-Thirring effect in general
relativity. In general, $N, b, K, \omega$ are functions of both
$r$ and $\theta$. $K(r,\theta)$ is a positive, nondecreasing
function that detemines the ``proper radial distance'' $R$
measured at $(r,\theta)$ from the origin.

\begin{figure}
\begin{center}
\includegraphics[width=6cm]{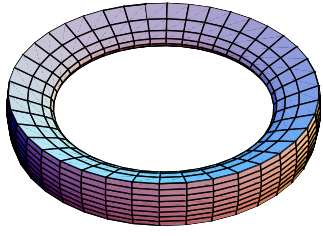}
\caption{Ergoregion for the dumbbell-like throat wormhole model.
It forms a tube.}
\end{center}
\end{figure}

\begin{figure}
\begin{center}
\includegraphics[width=4cm]{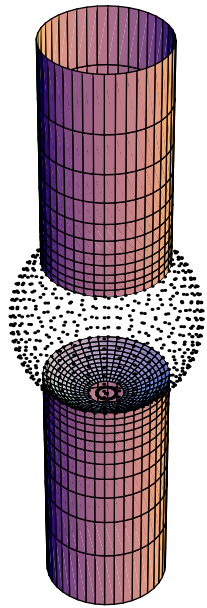}
\caption{Ergoregion of the rigid rotating wormhole model with
constant angular velocity. The dotted region is the wormhole
throat.}
\end{center}
\end{figure}

\begin{figure}
\begin{center}
\includegraphics[width=6cm]{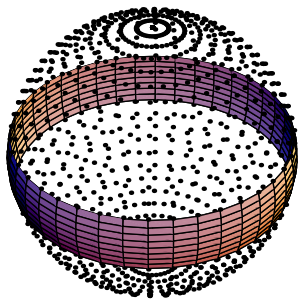}
\caption{Ergoregion of the rigid rotating wormhole model with
$\Omega=2a/r^3$. The dotted region is the wormhole throat. The
ergoregion also forms a tube.}
\end{center}
\end{figure}

It is regular on the symmetry axis $\theta = 0, \pi$ and two
identical, asymptotically flat regions joined together at throat
$r=b$. There are no event horizons or curvature singularities. The
off-diagonal component of the stress-energy tensor is $T_{t\phi}$
that is interpreted as the rotation of the matter distribution.

It clearly reduces to the MT metric in the limit of zero rotation
and spherical symmetry: \be N(r,\theta) \rightarrow
e^{\Lambda(r)}, \quad b(r,\theta) \rightarrow b(r), \quad
K(r,\theta) \rightarrow 1, \quad \omega(r,\theta) \rightarrow 0.
\ee

We also require that the metric should be asymptotically flat, in
which case \be N \rightarrow 1, \quad \f{b}{r} \rightarrow 0,
\quad K \rightarrow 1, \quad \omega \rightarrow 0 \ee as
$r\rightarrow \infty. $ Thus, $r$ is asymptotically the proper
radial distance. In particular, if \be \Omega =
\f{2a}{r^3}+O\left(\f{1}{r^4}\right), \ee then by changing to
Cartesian coordinates, it can be checked that $a$ is the total
angular momentum of the wormhole.

Afterwards, the source of the rotating wormhole was considered by
Bergliaffa and Hibberd\cite{BH00b}. They found the constraints on
the Einstein tensor that arise from the matter used as a source of
Einstein's equation for a generally axially symmetric and
stationary spacetime. The constraints are \be G_{11}-G_{22}=0,
\quad G_{12}=0, \quad G_{03}^2 = (G_{00}+G_{22})(G_{11}-G_{33})\ee
and \be G_{00}+G_{33} \ge 2G_{03}, \quad
\left(\f{u_0}{u_3}\right)^2
> 1, \ee where $u_\mu$ is the four-velocity of the
fluid model of the stress-energy tensor.

\section{Examples of rotating wormhole}

As the first example, Teo\cite{T98} took the gravitational
potential of rotating wormhole metric to be  \be
N=K=1+\f{(4a\cos\theta)^2}{r}, \quad b=1, \quad \omega =
\f{2a}{r^3}. \ee Note that $a$ is the angular momentum of the
resulting wormhole. Proper radius $R=1+(4a\cos\theta)^2$ is at
throat $r=1$ and the proper distance is \be l = \pm [
\sqrt{r(r-1)}+ \ln (\sqrt{r} + \sqrt{r-1})]. \ee

If the rotation of the wormhole is sufficiently fast, $g_{tt}$
becomes positive in some region outside the throat, indicating the
presence of an ergoregions where particles can no longer remain
stationary with respect to infinity. This occurs when $r^2 =
|2a\sin\theta|> 1$ when $|a|> 1/2$. The ergoregions does not
completely surround the throat, but forms a ``tube'' around the
equatorial region instead of ergosphere in Kerr metric. The plot
is in Figs. 1 and 2 which shows the cross-section of the throat
and the ergoregion.

As the second example, we just think about the rigid rotating
wormhole, i.e., \be N=K=1, \quad b=\f{b_0^2}{r}, \quad \Omega =
\mbox{const.} \ee such that the metric should be \be ds^2 = - dt^2
+ \f{dr^2}{1-\f{b(r)}{r}} + r^2 d\theta^2 + r^2\sin^2\theta [d\phi
- \Omega dt]^2. \label{eq:toy} \ee Without the angular velocity,
the spacetime is just the simplest wormhole model used in former
papers\cite{K00,KL01,KK98}. Here, the throat is $r=b_0$ and the
shape of the throat is a sphere. In this case, the ergoregion is
when $ r = 1/|\Omega \sin\theta|$. When $b_0\Omega > 1$, the
ergoregion is cut by throat as in Fig.~3, while the cylinder shape
ergoregion cover the whole wormhole throat, when $b_0\Omega \le
1$.

When the angular velocity is the same as first example, i.e.,
$\Omega=2a/r^3$ and when $r > b_0$, the ergoregion is tube type
for large angular momentum as shown in Fig.~4. It is similar to
Fig.~2.

\section{Perturbation of the wormhole}

\subsection{Scalar perturbation of non-rotating wormhole}

The spacetime metric for static uncharged wormhole is given as \be
ds^2 = -e^{2\Lambda(r)}dt^2 + \f{dr^2}{1-b(r)/r} + r^2
(d\theta^2+\sin^2\theta d\phi^2), \label{eq:mtwormhole} \ee where
$\Lambda(r)$ is the lapse function and $b(r)$ is the wormhole
shape function. They are assumed to be dependent on $r$ only for
static case.

The wave equation of the minimally coupled massless scalar field
is given by \be \nabla^\mu\nabla_\mu\Phi =\f{1}{\sqrt{-g}}
\partial_\mu
(\sqrt{-g} g^{\mu\nu}\partial_\nu \Phi ) = 0. \ee In spherically
symmetric space-time, the scalar field can be separated by
variables, \be \Phi_{lm} = Y_{lm}(\theta, \phi)\f{u_l(r,t)}{r},
\label{eq:def1} \ee where $Y_{lm}(\theta, \phi)$ is the spherical
harmonics and $l$ is the quantum angular momentum.

If the time dependence of the wave is harmonic as $u_l(r,t) =
\hat{u}_l(r,\omega)e^{-i\omega t} $, the equation becomes \be
\left( \f{d^2}{dr^2_*} + \omega^2 - V_l(r)
\right)\hat{u}_l(r,\omega) = 0, \ee where the potential is \Be
V_l(r)&=&\f{L^2}{r^2}e^{2\Lambda} + \f{1}{r}e^\Lambda
\sqrt{1-\f{b}{r}}\f{\partial}{\partial r}\left(e^\Lambda
\sqrt{1-\f{b}{r}}\right) \nonumber \\
&=& e^{2\Lambda}\left[ \f{l(l+1)}{r^2} - \f{b'r-b}{2r^3} +
\f{1}{r}\left(1-\f{b}{r}\right)\Lambda' \right]\nonumber\\ \Ee and
the proper distance $r_*$ has the following relation to $r$: \be
\f{\partial}{\partial r_*} = e^\Lambda r^2 \sqrt{1-\f{b}{r}}
\f{\partial}{\partial r}. \ee Here, $L^2=l(l+1)$ is the square of
the angular momentum. It is just the Schr\"odinger equation with
energy $\omega^2$ and potential $V_l(r)$. When $e^{2\Lambda}$ is
finite, $ V_l $ approaches zero as $r \rightarrow \infty $, which
means that the solution has the form of the plane wave $ \hat{u}_l
\sim e^{\pm i \omega r_*}$ asymptotically.

As the simplest example for this problem, we consider the special
case $ (\Lambda = 0, b = b_0^2/r)$ as usual, the potential should
be  \be V_l = \f{l(l+1)}{r^2} + \f{b_0^2}{r^4}, \label{eq:pot} \ee
where the proper distance $r_*$ is given by \be r_* = \int
\f{1}{\sqrt{1-b_0^2/r^2}}dr = \sqrt{r^2-b_0^2}. \label{eq:prop1}
\ee

\subsection{Scalar perturbation of rotating wormhole}

For the rotating case, we also only think the scalar field case in
this paper. By using $\sqrt{-g} = r^2K^2NL \sin\theta$, where $L^2
= 1/(1-b/r)$, the scalar wave equation becomes \be
\nabla_\mu\nabla^\mu\Phi = \f{1}{\sqrt{-g}}\partial_\mu (
g^{\mu\nu}\sqrt{-g}\partial_\nu \Phi) = 0. \ee In this case we try
to separate variables with $\Phi =
R(r)\Theta(\theta)e^{im\phi}e^{i\omega t}$. Then the wave equation
in spacetime of metric eq.~(\ref{eq:rot}) becomes \be
\f{\omega^2}{N^2} + \f{2\Omega}{N^2}\omega m - m^2 \left(
\f{1}{r^2K^2\sin^2\theta} - \f{\Omega^2}{N^2} \right) +
\f{1}{r^2K^2NL}\f{1}{R}\f{d}{dr}\left(
\f{r^2K^2N}{L}\f{d}{dr}\right) R   +
\f{1}{r^2K^2NL\sin\theta}\f{1}{\Theta}\f{d}{d\theta}\left( NL
\sin\theta \f{d}{d\theta}\right)\Theta = 0. \ee It is very hard to
separate variables, when $N, K, L$, and $\Omega$ are function of
$r$ and $\theta$.

To make the problem simple, we adapt the toy model as $N=K=1,
b(r)=\f{b_0^2}{r}$, and $\Omega(r)$ such that the metric of the
model spacetime is eq.~(\ref{eq:toy})\be ds^2 = - dt^2 +
\f{dr^2}{1-\f{b(r)}{r}} + r^2 d\theta^2 + r^2\sin\theta [d\phi^2 -
\Omega(r) dt]^2 \ee This is the just the rigid rotation of the
simplest model when $\Omega$ is constant. This model means that
the wormhole does not change its shape under rotation with angular
velocity $\Omega$. In both cases of constant $\Omega$ and
$r$-dependent $\Omega(r)$, the wave equation can be simply
separated as \be \f{1}{\sin\theta}\f{d}{d\theta}\left(\sin\theta
\f{d}{d\theta} \right)\Theta - \left( \f{m^2}{\sin^2\theta} -
\lambda_{lm} \right)\Theta = 0 \ee and \be \f{d^2u}{dr^2_*} +
V_l(\omega, m, r, \Omega)u = 0, \ee where $R(r) = \f{u(r)}{r}.$
The proper distance $r_*$ is defined as \be \f{d}{dr_*} =
\sqrt{1-\f{b(r)}{r}}\f{d}{dr} \ee which is the same as the
ron-rotating case\cite{K00}. The potential is \be V_l =
(\omega+\Omega m )^2 - \f{\lambda_{lm}}{r^2} - \f{2b_0^2}{r^4}.
\ee The separation constant $\lambda_{lm}$ becomes $l(l+1)$ in the
limit of $\omega = 0$.  The solution to $\Theta(\theta) =
S_{ml}(\theta,0)$ that is a spheroidal wave function in this case
becomes the spherical harmonics when $\omega\Omega=0$.

When $r \rightarrow \infty $ in constant $\Omega$ case, the
equation becomes \be \f{d^2u}{dr_*^2} + (\omega+\Omega m)^2u
\simeq 0 \ee which means \be u \sim e^{\pm i(\omega+\Omega m)
r_*}. \ee  When $\Omega=\f{2a}{r^3}$, the equation in the limit of
large $r$ becomes \be \f{d^2u}{dr_*^2} + \omega^2u \simeq 0 \ee
and the solutions are \be u \sim e^{\pm i\omega r_*}. \ee They are
same asymptotic solutions as those of non-rotating wormhole.

\section{Discussion}

We examined the rotating wormhole and found the Schr\"odinger type
equation for scalar perturbation in special model of rotating
wormhole. The result will be useful for considering the
gravitational wave problem by the wormhole and exotic matter.
Later, we will extend the problem into the gravitational
perturbation like the non-rotating case\cite{K04} to see the more
realistic situations.

\acknowledgments

This work was supported by grant No. R01-2000-000-00015-0 from the
Korea Science and Engineering Foundation.

\end{document}